 \documentclass[final,5p,times]{elsarticle}

\usepackage{amssymb}

\usepackage{graphicx}
\usepackage{hyperref}
\usepackage{amsmath}
\usepackage{booktabs}
\usepackage{threeparttable}
\biboptions{sort&compress}
\newcommand{\figref}[1]{Fig.~\ref{#1}}

\newcommand{\tabref}[1]{Tab.~\ref{#1}}

\journal{Physics Letters B}

\begin{document}

\begin{frontmatter}



\title{Electromagnetic moments of the antimony isotopes $^{112-133}$Sb}

\author[1,2]{S. Lechner\corref{mycorrespondingauthor}}
\cortext[mycorrespondingauthor]{Corresponding authors:}
\ead{simon.lechner@cern.ch}
\author[3,4,5,6]{T. Miyagi}
\author[7,8]{Z. Y. Xu}
\author[9,1]{M. L. Bissell}
\author[5] {K. Blaum}
\author[10]{B. Cheal}
\author[10]{C. S. Devlin}
\author[1]{R. F. Garcia Ruiz\fnref{rfgr}}
\fntext[rfgr]{Present address: Massachusetts Institute of Technology, Cambridge, MA, USA}
\author[13]{J. S. M. Ginges}
\author[1,5]{H. Heylen}
\author[6]{J. D. Holt}
\author[3,7]{P. Imgram}
\author[7]{A. Kanellakopoulos\fnref{AKa}}
\fntext[AKa]{Present address:HEPIA Geneva, HES-SO, 1202 Geneva, Switzerland}
\author[1,7]{\'{A}. Koszor\'{u}s}
\author[1,6,14]{S. Malbrunot-Ettenauer\corref{mycorrespondingauthor}}
\ead{sette@triumf.ca}
\author[5,15]{R. Neugart}
\author[1,7]{G. Neyens}
\author[3]{W. N\"ortersh\"auser}
\author[1,16,5]{P. Plattner}
\author[5,17]{L. V. Rodr\'iguez}
\author[13]{G. Sanamyan}
\author[19]{S. R. Stroberg\fnref{SRS}}
\fntext[SRS]{Present address: University of Notre Dame, Notre Dame IN, USA}
\author[20,21]{Y. Utsuno}
\author[22]{X. F. Yang}
\author[17]{D. T. Yordanov}

\address[1]{Experimental Physics Department, CERN, CH-1211 Geneva 23, Switzerland}
\address[2]{Technische Universität Wien, AT-1040 Wien, Austria}
\address[3]{Institut für Kernphysik, Department of Physics, TU Darmstadt, D-64289 Darmstadt, Germany}
\address[4]{ExtreMe Matter Institute EMMI, GSI Helmholtzzentrum für Schwerionenforschung GmbH, 64291 Darmstadt, Germany}
\address[5]{Max-Planck-Institut für Kernphysik, Saupfercheckweg 1, 69117 Heidelberg, Germany}
\address[6]{TRIUMF, Vancouver, British Columbia V6T 2A3, Canada}
\address[7]{Instituut voor Kern- en Stalingsfysica, KU Leuven, B-3001, Leuven, Belgium}
\address[8]{Department of Physics and Astronomy, University of Tennessee, 37996 Knoxville, TN, USA}
\address[9]{School of Physics and Astronomy, University of Manchester, Manchester, M13 9PL, United Kingdom}
\address[10]{Oliver Lodge Laboratory, University of Liverpool, Liverpool, L69 7ZE, United Kingdom}
\address[13]{School of Mathematics and Physics, The University of Queensland, Brisbane QLD 4072, Australia}
\address[14]{Department of Physics, University of Toronto, 60 St. George St., Toronto, Ontario, Canada}
\address[15]{Institut für Kernchemie, Universität Mainz, D-55128 Mainz, Germany}
\address[16]{Universität Innsbruck, AT-6020 Innsbruck, Austria}
\address[17]{Université Paris-Saclay, CNRS-IN2P3, IJCLab, 91405 Orsay, France}
\address[19]{Department of Physics, University of Washington, Seattle, WA, USA}
\address[20]{Center for Nuclear Study, The University of Tokyo, 7-3-1 Hongo, Bunkyo-ku, Tokyo 113-0033, Japan}
\address[21]{Advanced Science Research Center, Japan Atomic Energy Agency, Tokai, Ibaraki 319-1195, Japan}
\address[22]{School of Physics and State Key Laboratory of Nuclear Physics and Technology, Peking University, Beijing 10871, China}
\address[17]{Université Paris-Saclay, CNRS-IN2P3, IJCLab, 91405 Orsay, France}

\begin{abstract}
Nuclear moments of the antimony isotopes $^{113-133}$Sb are measured by collinear laser spectroscopy and used to benchmark phenomenological shell-model and \textit{ab initio} calculations in the valence-space in-medium similarity renormalization group (VS-IMSRG).
The shell-model calculations reproduce the electromagnetic moments over all Sb isotopes when suitable effective $g$-factors and charges are employed. Good agreement is achieved by VS-IMSRG for magnetic moments on the neutron-deficient side for both odd-even and odd-odd Sb isotopes while its results deviate from experiment on the neutron-rich side. When the same effective $g$-factors are used, VS-IMSRG agrees with experiment nearly as well as the shell model. Hence, the wave functions are very similar in both approaches and missing contributions to the M1 operator are identified as the cause of the discrepancy of VS-IMSRG with experiment. Electric quadrupole moments remain more challenging for VS-IMSRG.
\end{abstract}

\begin{keyword}


Collinear laser spectroscopy,
Electromagnetic moments, 
Ab-initio calculation

\end{keyword}

\end{frontmatter}

\section{Introduction}

Electromagnetic moments are intriguing observables of atomic nuclei which played a major role in the development of the nuclear shell model in the last century \cite{Jensen1949,Mayer1950_1,Mayer1950_2}.
While nuclear magnetic dipole moments are sensitive to the single-particle behaviour of unpaired valence nucleons, nuclear electric quadrupole moments characterize the non-sphericity, i.e.\ deformation, of the nuclear charge distribution \cite{Neyens2003,deGroote2022}.

Modern, phenomenological formulations of the shell model have been very successful in describing atomic nuclei \cite{Otsuka2018}.
In the case of electromagnetic moments, this is achieved by introducing effective $g$-factors and charges in the respective operators \cite{Neyens2003}.
Their purpose is to provide an ad hoc representation of higher-order physics such as core-polarization effects or meson-exchange currents, which is not explicitly modelled in the shell-model calculations.
As a consequence, they need to be adjusted for different mass regions and model spaces \cite{Bogner2002,Coraggio2019}.
One major effort in the theoretical description of electromagnetic moments is hence to provide the underlying, microscopic foundation of effective $g$-factors and charges.

The rapid advances in nuclear many-body methods over the last decade \cite{hagen2015,Morris2018,gysbers2019,Hergert2020,Arthuis2020,Stroberg2021,Soma2021} have raised the prospect that, in the long term, nuclear phenomena could be consistently described all across the nuclear chart when employing nuclear forces which are constructed in chiral effective field theory and rooted in the symmetries of quantum chromodynamics \cite{Epelbaum2009,MACHLEIDT2011,Machleidt2020,Hammer2020}.
Such \textit{ab initio} approaches additionally offer the machinery to systematically evolve the respective nuclear operators such that effective $g$-factors and charges should in principle be derivable from first principles.
While \textit{ab initio} calculations have made significant progress also for electromagnetic moments \cite{Parzuchowski2017,Klose2019,Heylen2021,Vernon2022}, remaining systematic discrepancies to experimental data demand further theoretical developments.

This letter reports spins and electromagnetic moments of the antimony ($Z=51$) isotopes $^{113-133}$Sb measured via high-resolution laser spectroscopy.
Due to a single valence proton outside the $Z=50$ proton-shell closure, the nuclear magnetic moments of odd-even Sb isotopes are expected to follow single-particle behaviour and can be well described by conventional shell-model considerations.
In this study, the experimental data is compared to large-scale shell-model calculations and the valence-space in-medium similarity renormalization group (VS-IMSRG) \cite{Hergert2016,Stroberg2019}, a leading \textit{ab initio} method.
While individual magnetic moments of Sb isotopes have been calculated in Ref.~\cite{Stone1997,Borzov2008,Achakovskiy2014}, the theoretical approaches employed in the present work utilize the same valence-space diagonalization.
Hence, the full comparison of these two theoretical models in respect to the experimental data as well as the artificial use of the shell model’s effective $g$-factors and charges in VS-IMSRG allows an investigation of the operator construction and renormalization within VS-IMSRG separated from the obtained nuclear wave functions.
Conducting such a study in Sb isotopes is of special interest when \textit{ab initio} theory has just begun to expand its reach beyond the shell closure $Z=50$ and into heavier systems.

\section{Experiment and Analysis}
Details of the experiment have already been outlined in \\ Ref.~\cite{Lechner2021}.
In brief, radioactive antimony (Sb) isotopes were produced at the radioactive ion beam facility ISOLDE-CERN \cite{Catherall_2017}.
Neutron-deficient Sb isotopes were generated directly from proton collisions with an uranium carbide target.
For neutron-rich isotopes, a neutron converter was utilized to enable neutron-induced fission of uranium which suppressed contamination.
Following selective resonant laser ionization \cite{Fedosseev_2017} and mass separation, cooled ion bunches were prepared in a radio-frequency quadrupole cooler and buncher \cite{FRANBERG2008}.
These ion bunches were transferred at energies of around 50\,keV to the COLLAPS beam line, where they were collinearly overlapped with a narrowband continuous-wave laser beam.
After the ions were neutralized in a charge-exchange cell, photons emitted from resonantly excited Sb atoms were detected by four photomultiplier tubes.
Scanning the laser frequency was achieved by altering the floating potential of the charge-exchange cell and thus, the atoms' kinetic energy.
A frequency-quadrupled Ti:Sa laser at 217\,nm was driving the atomic transition $5s^25p^3\ ^4\text{S}_{3/2} \rightarrow 5s^25p^26s\ ^4\text{P}_{3/2}$.

\begin{figure}
    \begin{center}
        \includegraphics[width=0.49\textwidth]{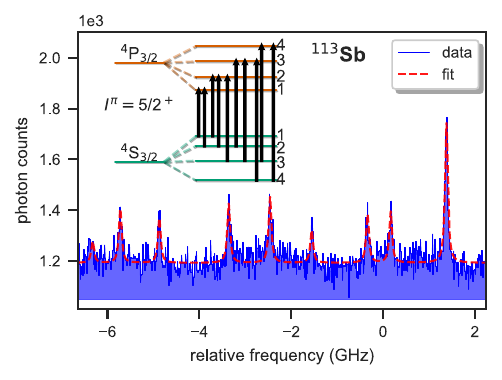}
    \end{center}
    \caption{Hyperfine spectrum of $^{113}$Sb in the $5s^25p^3\ ^4\text{S}_{3/2} \rightarrow 5s^25p^26s\ ^4\text{P}_{3/2}$ transition including a scheme of the hyperfine splitting.
    The frequency is given as an offset from the transition frequency (45945.340(5)\,cm$^{-1}$ \cite{NIST_ASD}).}
    \label{fig:spectrum}
\end{figure}

In this manner, hyperfine spectra of $^{113-133}$Sb ($N = 61 - 82$), including long-lived isomeric states, were recorded.
A spectrum of $^{113}$Sb is shown in \figref{fig:spectrum}.
The hyperfine parameter $A$ and $B$ were obtained for the lower and upper level of the transition by fitting the hyperfine spectra with the SATLAS package \cite{GINS2018}.
\begin{table*}[ht]
\centering
\caption{Hyperfine parameters $A$ and $B$ from the lower ($5s^25p^3\ ^4\text{S}_{3/2}$) and upper ($5s^25p^26s\ ^4\text{P}_{3/2}$) level for all measured Sb isotopes in comparison to literature.
Note that $B$\textsubscript{lower} of $^{121,123}$Sb was fixed to the given literature values, while $B$\textsubscript{lower} for all other isotopes was derived from $B$\textsubscript{upper} using the $B$-ratio $B_{\text{upper}}/B_{\text{lower}} = 128.6(6)$, which was fixed in the fit of the hyperfine spectra.
$A$\textsubscript{lower} and $B$\textsubscript{upper} of $^{123}$Sb slightly changed compared to the previous publication of this work \cite{Lechner2021} based on re-analysis.
$^{121,133}$Sb are not affected by the re-analysis and the same values from Ref.~\cite{Lechner2021} are given for completeness.
Isomers are indicated by $A$\textsuperscript{m}.
}
 \vspace{1mm}
\label{tab:hyperfine_parameter}
\begin{tabular}{@{}llllllllll@{}}
\hline
\hline
            &           & \multicolumn{2}{l}{$A_{\text{lower}}$ / MHz}     & \multicolumn{2}{l}{$A_{\text{upper}}$ / MHz} & \multicolumn{2}{l}{$B_{\text{lower}}$ / MHz}  & \multicolumn{2}{l}{$B_{\text{upper}}$ / MHz} \\ \cmidrule(l){3-10} 
$A$ & $I^{\pi}$ & Exp.       & Lit.                            & Exp.      & Lit.                         & Exp.       & Lit.                         & Exp.        & Lit.                       \\ \midrule
113 & $5/2^+$ & -303.0(6) &  & 528.9(6) &  & -3.85(3) &  & -494(3) &  \\
115 & $5/2^+$ & -307.3(3) & -307.68(19) \cite{Jackson1968} & 539.2(2) &  & -3.71(3) & -3.7(5) \cite{Jackson1968} & -478(4) &  \\
116\textsuperscript{m} & $8^-$ & -64.49(7) &  & 112.94(6) &  & -6.22(3) &  & -800.0(14) &  \\
117 & $5/2^+$ & -311.9(2) & -305(5) \cite{EKSTROM1974} & 546.5(2) &  & -3.56(2) & 3(21) \cite{EKSTROM1974} & -458.3(13) &  \\
118\textsuperscript{m} & $8^-$ & -65.83(4) &  & 115.31(4) &  & -5.28(3) &  & -678.8(8) &  \\
119 & $5/2^+$ & -307.0(2) & -307.16(6) \cite{Jackson1968} & 537.9(2) &  & -3.64(2) & -3.8(4) \cite{Jackson1968} & -468.1(10) &  \\
120\textsuperscript{m} & $8^-$ & -65.38(6) &  & 114.67(4) &  & -4.69(2) &  & -603.1(11) &  \\
121 & $5/2^+$ & -299.0(3) & -299.034(4) \cite{Fernando1960} & 523.8(5) & 519(6) \cite{Hassini88} &  & -3.68(2) \cite{Fernando1960} & -471(2) & -480(15) \cite{Hassini88} \\
122\textsuperscript{m} & $(8)^-$ & -39.8(2) &  & 70.3(2) &  & -4.28(5) &  & -550(6) &  \\
123 & $7/2^+$ & -162.58(8) & -162.451(3) \cite{Fernando1960} & 285.20(8) & 282(4) \cite{Sobolewski2016} &  & -4.67(3) \cite{Fernando1960} & -602.7(7) &  \\
124\textsuperscript{m} & $8^-$ & -32.80(9) &  & 57.38(10) &  & -2.88(2) &  & -370(2) &  \\
125 & $7/2^+$ & -168.2(2) &  & 294.7(2) &  & -4.30(3) &  & -553(2) &  \\
126 & $8^-$ & -31.51(7) &  & 55.54(7) &  & -1.847(13) &  & -237.5(13) &  \\
127 & $7/2^+$ & -175.7(2) &  & 307.9(2) &  & -3.90(2) &  & -501(2) &  \\
128 & $8^-$ & -32.79(13) &  & 57.7(2) &  & -0.962(11) &  & -123.7(13) &  \\
129 & $7/2^+$ & -183.1(3) &  & 321.3(3) &  & -3.37(3) &  & -433(4) &  \\
130 & $8^-$ & -34.97(8) &  & 61.78(8) &  & -0.220(13) &  & -28(2) &  \\
131 & $7/2^+$ & -189.7(2) &  & 333.2(2) &  & -2.72(3) &  & -350(3) &  \\
132\textsuperscript{m} & $(8^-)$ & -38.15(15) &  & 66.9(2) &  & 0.29(3) &  & 38(3) &  \\
133 & $7/2^+$ & -196.1(2) &  & 343.7(2) &  & -2.06(2) &  & -265(2) &  \\
\hline
\end{tabular}
\end{table*}
Table~\ref{tab:hyperfine_parameter} lists all measured hyperfine parameters $A$ and $B$ in comparison to literature.
$B$\textsubscript{lower} was too small to be properly extracted from the data.
Hence, the ratio of $B$\textsubscript{upper}/$B$\textsubscript{lower} was fixed to 128.6(6) for the analysis \cite{Lechner2021}.
This ratio was obtained by constraining $B$\textsubscript{lower} of $^{121,123}$Sb to the literature values from Ref.~\cite{Fernando1960} during the fit and then taking the weighted average of both isotopes.

For several isotopes the spins had been only tentatively assigned in the literature \cite{nndc}.
Except for the $I^{\pi}=8^-$ states in $^{122}$Sb and $^{132}$Sb, all observed states were firmly assigned by using different spins for fitting the spectra and determining the minimum of $\chi^2$ (see Refs.~\cite{Lechner2021,Lechner_thesis2021} for details).
Due to systematics from other odd-odd isotopes with an $8^-$ state and the tentative assignment in Ref.~\cite{nndc}, $I^{\pi}=(8^-$) will be used in the following for $^{122}$Sb and $^{132}$Sb.

\section{Calculation of the hyperfine anomaly}
To extract the magnetic moments $\mu$ from the measured hyperfine parameters $A$ with the help of a reference isotope with known $\mu$, the hyperfine anomaly~\cite{Buettgenbach1984,Persson2023} has to be taken into account; see below in Sec.~\ref{sec:results}.
The relevant differential anomaly $\Delta$ may be expressed in terms of the difference in the Breit-Rosenthal (BR) and the Bohr-Weisskopf (BW) effects for the considered isotopes~\cite{Buettgenbach1984,Persson2023},
\begin{equation}
\Delta \approx \delta_{\rm ref} - \delta+\epsilon_{\rm ref} - \epsilon\, ,
\end{equation}
where the BR contribution $\delta$ to the hyperfine structure arises from the finite distribution of nuclear charge and the BW effect $\epsilon$ from the finite distribution of nuclear magnetisation.
The former is taken into account through the modelling of the nuclear charge distribution in the relativistic Hamiltonian for the atomic electrons; we use a Fermi distribution with a skin thickness of 2.3\,fm and rms charge radii estimated from an empirical interpolation formula~\cite{Johnson1985}. 
The latter is considered through a modification to the hyperfine operator; to obtain this, we use the nuclear single-particle model with the unpaired nucleon wave function found in the Woods-Saxon potential and spin-orbit interaction included~\cite{Bohr1950,Bohr1951,Shabaev1997}. The orbital $g$-factor contribution is taken to be that of a free nucleon, and the spin $g$-factor contribution $g_s$ is obtained from the (uncorrected) measured value for the magnetic moment, using a similar procedure as in Ref.~\cite{Shabaev1997}. For even isotopes with a proton and neutron contribution to the BW effect, the $g_s$ contribution for the proton is found from a neighbouring odd isotope, and that for the neutron from the considered isotope~\cite{Shabaev1995}.  
For isotopes with different nuclear spins, the difference in BW effects typically dominates; however, when the spins are the same the differential BW effect may be very small, and accurate account of the differential BR effect may be important.

To evaluate the differential anomalies $\Delta$ in the atomic state $5s^25p^3\, ^4{\rm S}_{3/2}$ we perform the calculations in two stages.
First, we determine an isotope-independent electronic screening factor~\cite{Roberts2022} which relates the effect in the many-electron atom to that in the hydrogen-like ion, $\Delta \approx x_{\rm scr}\Delta^{\rm H-like}$, where we take the screening to be the same for the BR and BW effects.
Then all calculations of the nuclear structure effects may be carried out for the simple case of the corresponding hydrogen-like ion.
We obtained the screening factor $x_{\rm scr}=0.848(24)$ through a semi-empirical approach, by comparing the measured differential anomaly $^{123}\Delta^{121}$~\cite{Persson2023} with the anomaly calculated for the H-like system in the nuclear single-particle model.
We checked this result with {\it ab initio} atomic many-body calculations performed with the GRASP many-body suite of codes~\cite{GRASP}, from which we obtained the factor $0.909$, with a larger uncertainty than the semi-empirical value. 
The uncertainty assigned to $\Delta$ comes from the uncertainty in the screening factor and the (dominating) uncertainty in the nuclear single-particle model and in the modelling of the nuclear charge distribution. 
We estimate the uncertainty for evaluation of the BW effect for odd isotopes to be 10\%, and 20\% for even isotopes. We consider this to be reasonable based on the good performance of the calculated differential anomalies for nearby atoms of silver and cesium, where experimental data for several isotopes are available~\cite{Roberts2021,Sanamyan2023,Persson2023}, and because antimony lies in the region where the single-particle model works well (one proton away from magic).  
The calculated differential anomalies are presented in \tabref{tab:moments}. 

\section{Results}\label{sec:results}
\begin{table*}[ht!]
\centering
\caption{Nuclear magnetic dipole and electric quadrupole moments of Sb isotopes obtained in this work in comparison to literature.
The differential hyperfine anomaly $\Delta$ with respect to $^{123}$Sb is included in the evaluations of the magnetic moments (see text for details). 
Isomers are indicated by $A$\textsuperscript{m}.
All moments are derived by using $^{123}$Sb as reference.
Note that the numerical value of the magnetic moment of $^{133}$Sb is slightly different from the previous publication of this work \cite{Lechner2021}, since $\Delta$ was not included, while $A$\textsubscript{upper} was considered (see Ref.~\cite{Lechner2021}).
Quadrupole moments of $^{121,133}$Sb from Ref.~\cite{Lechner2021} are listed for completeness.
}
\vspace{1mm}
\label{tab:moments}
\begin{threeparttable}
\begin{tabular}{@{}lllllll@{}}
\hline
\hline
            &     &      & \multicolumn{2}{l}{$\mu$ / $\mu_N$}                                            & \multicolumn{2}{l}{Q / b}                          \\ \cmidrule(l){4-7} 
A & $I^{\pi}$ & $\Delta(\%)$ &Exp.              & Lit.                                                       & Exp.       & Lit.                                  \\ \midrule
113 & $5/2^+$ & 0.30(6) & $3.402(7)$ &  & -0.568(12) &  \\
115 & $5/2^+$ & 0.31(6) & $3.450(4)$ & 3.46(1) \cite{Jackson1968} & -0.548(12) & -0.546(75) \cite{Jackson1968} \\
116\textsuperscript{m} & $8^-$ & 0.09(6) & $2.312(3)$ & 2.59(22) \cite{Booth1993} & -0.92(2) &  \\
117 & $5/2^+$ & 0.31(6) & $3.502(3)$ & 3.42(6)\tnote{a} \cite{EKSTROM1974} & -0.526(11) & 0.2(12) \cite{EKSTROM1974} \\
118\textsuperscript{m} & $8^-$ & 0.10(6) & $2.360(2)$ & 2.32(4) \cite{CALLAGHAN1974,Stone2019} & -0.78(2) &  \\
119 & $5/2^+$ & 0.32(6) & $3.447(3)$ & 3.45(1) \cite{Jackson1968} & -0.537(11) & -0.561(60) \cite{Jackson1968} \\
120\textsuperscript{m} & $8^-$ & 0.11(6) & $2.344(3)$ & 2.34(3) \cite{CALLAGHAN1974,Stone2019} & -0.692(14) &  \\
121 & $5/2^+$ & 0.323(9)\tnote{b} & $3.358(4)$ & 3.3580(16) \cite{Proctor1951,Stone2019} & -0.541(11) & -0.543(11) \cite{Haiduke2006} \\
122\textsuperscript{m} & $(8)^-$ & -0.38(7) & $1.420(7)$ &  & -0.632(15) &  \\
123 & $7/2^+$ & 0 &  & 2.5457(12) \cite{Proctor1951,Stone2019} &  & -0.692(14) \cite{Haiduke2006} \\
124\textsuperscript{m} & $8^-$ & -0.59(11) & $1.168(3)$ &  & -0.425(9) &  \\
125 & $7/2^+$ & 0.03(3) & $2.637(3)$ & 2.63(4) \cite{CALLAGHAN1974} & -0.635(13) &  \\
126 & $8^-$ & -0.63(12) & $1.121(3)$ & 1.28(7) \cite{Krane1972} & -0.273(6) &  \\
127 & $7/2^+$ & 0.06(3) & $2.755(3)$ & 2.70(4) \cite{Lindroos1996,Stone2019}, 2.59(12) \cite{Krane1972} & -0.576(12) &  \\
128 & $8^-$ & -0.58(11) & $1.168(5)$ & 1.3(2) \cite{Krane1972} & -0.142(3) &  \\
129 & $7/2^+$ & 0.09(3) & $2.872(5)$ & 2.79(4) \cite{Stone1997,Stone2019} & -0.497(11) &  \\
130 & $8^-$ & -0.49(9) & $1.246(3)$ &  & -0.033(2) &  \\
131 & $7/2^+$ & 0.12(3) & $2.976(4)$ & 2.89(4) \cite{Stone1997,Stone2019} & -0.402(9) &  \\
132\textsuperscript{m} & $(8^-)$ & -0.39(7) & $1.361(5)$ &  & 0.043(4) &  \\
133 & $7/2^+$ & 0.14(3) & $3.077(4)$ & 3.00(4) \cite{Stone1997,Stone2019}, 3.070(2) \cite{Lechner2021} & -0.304(7) & \\
\hline
\end{tabular}
\begin{tablenotes}\footnotesize
\item[a] Re-evaluated in the present work by using the updated reference value from Ref.~\cite{Stone2019}.
\item[b] Empirical value for differential anomaly,  Ref.~\cite{Persson2023}.

\end{tablenotes}
\end{threeparttable}
\end{table*}

Nuclear magnetic dipole moments $\mu$ and electric quadrupole moments $Q$ were derived from the hyperfine parameters $A$\textsubscript{lower} and $B$\textsubscript{upper} with respect to the reference isotope $^{123}$Sb:
\begin{equation}
	\label{eq:magnetic_moment_hfa}
  \mu = \mu_{\text{ref}} \frac{AI}{A_{\text{ref}}I_{\text{ref}}} (1 + \Delta)
\end{equation}
and
\begin{equation}
	\label{eq:Q_moment_ref}
  Q = Q_{\text{ref}} \frac{B}{B_{\text{ref}}}\, ,
\end{equation}
where $\Delta$ is the differential hyperfine anomaly as calculated above. $\mu_\text{123Sb}$ was determined from the results of the nuclear magnetic resonance measurement in Ref.~\cite{Proctor1951} applying the diamagnetic correction from Ref.~\cite{Stone2019} estimated through systematic considerations.
$Q_\text{123Sb}$ was calculated by the advanced molecular method from the experimental nuclear quadrupole coupling constants in SbN and SbP molecules \cite{Haiduke2006}. Since the hyperfine anomaly calculations were performed for the lower atomic state $5s^25p^3\ ^4\text{S}_{3/2}$, only $A$\textsubscript{lower} was used in Eq.~\ref{eq:magnetic_moment_hfa} to determine the magnetic moments. Similarly for the quadrupole moments, only $B$\textsubscript{upper} was employed due to the fixed $B$-ratio.

All obtained electromagnetic moments are listed in comparison to literature in \tabref{tab:moments}. Although a large majority of the hyperfine parameters were measured for the first time in the present work, a series of electromagnetic moments of Sb isotopes had previously been determined by other experimental techniques.
Our studies of 5/2$^{+}$, 7/2$^{+}$, 8$^{-}$ states in Sb isotopes revealed 5 new magnetic moments and 15 new electric quadrupole moments and, with the exception of stable $^{121}$Sb, improved the experimental precision in all other cases, typically by about one order of magnitude.

Overall the agreement between this work and literature is reasonable, but a few discrepancies were identified for certain magnetic moments.
In particular, a deviation is apparent for values measured with nuclear magnetic resonance of oriented nuclei (NMR/ON), where a certain magnetic hyperfine field $B_{\text{hf}}$ in iron, taken from Ref.~\cite{Koi1972}, was used ($^{116\text{gs},127,129,131,133}$Sb).
On the other hand, experiments using the same technique but a different hyperfine field from Ref.~\cite{CALLAGHAN1974}, show good agreement with this work ($^{118\text{m},120\text{m},125}$Sb).
As already discussed in Refs.~\cite{Lechner2021,Lechner_thesis2021}, $B_{\text{hf}}$ from Ref.~\cite{CALLAGHAN1974} leads to consistent results for all isotopes and should therefore be used in the future.
Another disagreement is found in $^{126}$Sb, which was measured with low-temperature static nuclear orientation (NO/S).
NO/S depends on several input parameters, which might have introduced an unaccounted, systematic uncertainty.

\section{Methods employed in nuclear-structure calculations}

\subsection{Shell-Model Calculations} 
\label{sub:shell_model_calculations}

\begin{figure}
    \begin{center}
        \includegraphics[width=0.4\textwidth]{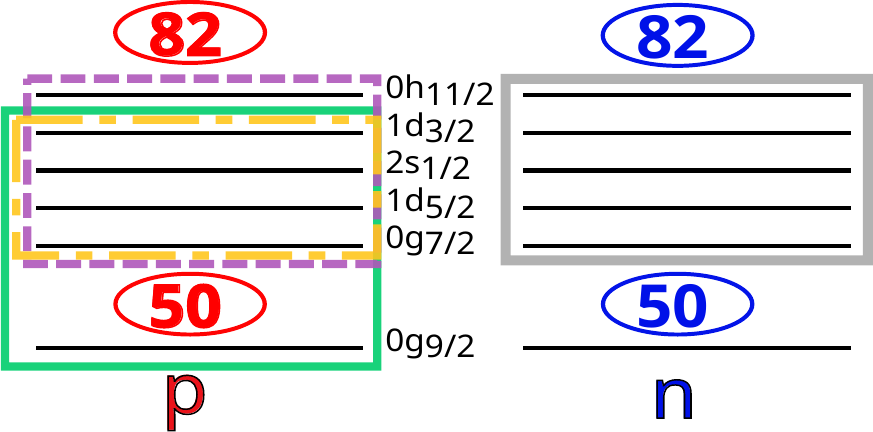}
    \end{center}
    \caption{Relevant orbitals for the Sb isotopic chain.
    All calculation methods include the full neutron $N=50-82$ valence space.
    For the shell-model calculations also the full proton $Z=50-82$ valence space (dashed) is used, while for the \textit{ab initio} calculations either the sdg7 (dash-dot) or sdg (full line) valence space is employed.}
    \label{fig:orbitals}
\end{figure}

Large-scale shell-model calculations are carried out with 
the valence shell consisting 
of the proton and neutron orbitals $\{0g_{7/2}, 1d_{5/2}, 2s_{1/2}, 1d_{3/2}, 0h_{11/2}\}$, see \figref{fig:orbitals}.
The effective interaction used here is the same as the one of 
Ref.~\cite{Utsuno2014}; the neutron-neutron 
and proton-neutron interactions are
taken from the SNBG3 \cite{Honma2008}.
Since Sb isotopes have only one valence proton, no two-body proton-proton force has to be considered.
The employed $V_{MU}$ contains central, two-body spin-orbit, and tensor forces \cite{Utsuno2012}.
Its central force is scaled by 0.84 to reproduce one-proton separation energies of Sb isotopes. 
As presented in Ref.~\cite{Utsuno2014}, the evolution of energy 
spacings among the $5/2^+_1$, $7/2^+_1$ and $11/2^-_1$ levels 
in Sb isotopes are well reproduced, in which the tensor force 
plays a crucial role. 
Here the shell-model Hamiltonian matrices constructed by this effective interaction 
are diagonalized by using the KSHELL code \cite{SHIMIZU2019}, and 
magnetic dipole moments and electric quadrupole moments are calculated
with the eigenvectors thus obtained.
The effective $g$ factors adopted are $(g_l^{\pi}, g_l^{\nu})=
(1.11, -0.02)$ and $g_{\text{s,eff}}=0.6g_{\text{s}}$
to reproduce the magnetic moments of $^{133}$Sb($7/2^+_1$),  
$^{131}$Sn($3/2^+_1$), and $^{131}$Sn($11/2^-_1$). 
The effective proton charge $e_{\pi}=1.6$ is determined from $^{133}$Sb, since there is only a single proton contribution to its quadrupole moment in the calculations due to the closed neutron shell at $N=82$.
The effective neutron charge $e_{\nu}=+1.05$ is obtained from a fit of all measured quadrupole moments of this work.

\subsection{\textit{Ab initio} Calculations} 
\label{sub:ab_initio_calculations}
The valence-space in-medium similarity renormalization\\ group (VS-IMSRG)~\cite{Stroberg2019} approach is used to compute the relevant observables.
This approach is based on the underlying nuclear interaction.
Our calculation of the electromagnetic moments begins with the nucleon-nucleon (NN) and three-nucleon (3N) 1.8/2.0 (EM) interaction~\cite{Hebeler2011}, expressed with the 13 major shell spherical harmonics-oscillator basis at the frequency of $12$\,MeV.
The 1.8/2.0 (EM) interaction is fitted to replicate only few-body data, while ground-state energies for medium- and heavy-mass nuclei~\cite{Simonis2017,Stroberg2021,Miyagi2022} are well reproduced.
Absolute charge radii~\cite{Simonis2017} are underestimated though.
We employ the novel storage scheme for 3N force matrix elements~\cite{Miyagi2022}, which greatly reduces the required memory and hence, makes converged calculations of the Sb isotopic chain feasible.
With a sufficiently large truncation of $E_{\rm 3max}=24$, all observables studied in this work are converged.
We include approximate 3N forces between valence-space nucleons with the ensemble normal ordering technique~\cite{Stroberg2017}, and truncate many-body operators at the two-body level, which is known as the IMSRG(2) approximation.
Effective valence-space electromagnetic operators relevant to this work are also decoupled with the same transformation as the Hamiltonian~\cite{Parzuchowski2017}, enabling us to include core-polarization effects in a non-perturbative way~\cite{Stroberg2019}.
In the current study, using the multi-shell approach of Ref.~\cite{Miyagi2020}, our valence space spans the proton $\{0g_{7/2}, 1d_{5/2}, 2s_{1/2}, 1d_{3/2}\}$ and neutron $\{0g_{7/2}, 1d_{5/2}, 2s_{1/2}, 1d_{3/2}, 0h_{11/2}\}$ orbitals above the $^{100}$Sn core, see \figref{fig:orbitals}.
For selected isotopes, additional calculations were performed including the proton orbital $0g_{9/2}$.
The VS-IMSRG decoupling is performed with the imsrg++ code~\cite{imsrg++}, and the full valence-space diagonalizations and corresponding one- and two-body transition-density calculations are done with the KSHELL code~\cite{SHIMIZU2019}.


\section{Discussion}
\subsection{Magnetic Moments}

\begin{figure*}
    \begin{center}
        \includegraphics[width=0.9\textwidth]{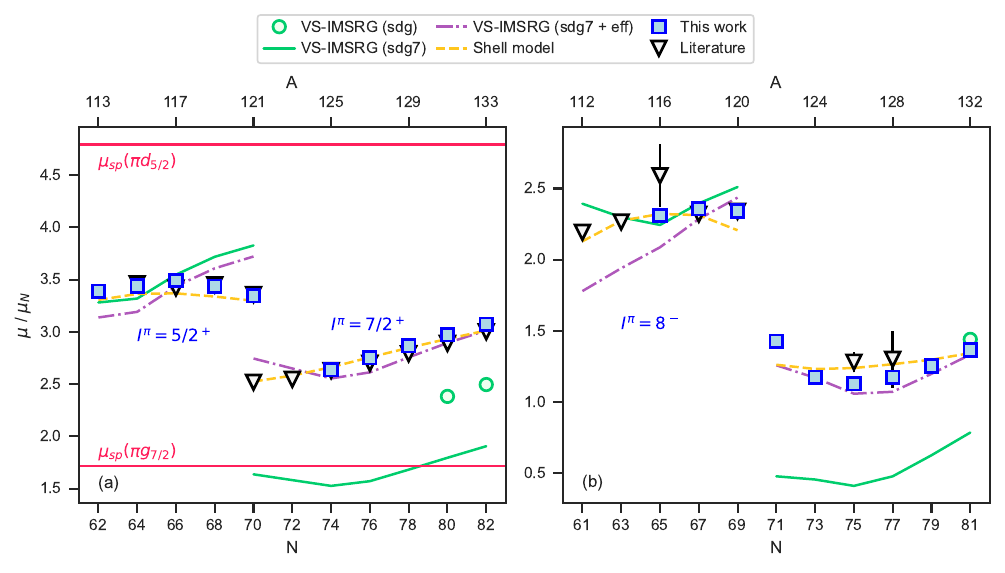}
    \end{center}
    \caption{Magnetic moments of (a) odd-even and (b) odd-odd $^{112-133}$Sb isotopes in comparison to shell-model and \textit{ab initio} calculations within the VS-IMSRG framework and the EM1.8/2.0 nuclear interaction. 
    $\mu_\text{sp}$ indicates the single-particle value for the respective orbital.
    (sdg) includes the proton orbitals $\{0g_{9/2}, 0g_{7/2}, 1d_{5/2}, 2s_{1/2}, 1d_{3/2}\}$ and (sdg7) the same except $0g_{9/2}$, see \figref{fig:orbitals}.
    ``eff'' means that the effective $g$-factors $g_{\text{s,eff}}=0.6g_{\text{s}}$, $g_{\text{l,}\pi}=1.11$ and $g_{\text{l},\nu}=-0.02$ from the shell-model calculations were used instead of the renormalization of the operator.
    Literature values are taken from Refs.~\cite{Proctor1951,Fernando1960,Jackson1968,Krane1972,EKSTROM1974,CALLAGHAN1974,Ketel1976,Langouche1976,Booth1993,Lindroos1996,Stone1997,Stone2019}.
    Note that the second literature value of $^{121}$Sb is the first excited $7/2^+$ state (not observed in this work due to its short half-life).
}
    \label{fig:mu}
\end{figure*}

In the single-particle picture, up to $^{121}$Sb, the single proton of antimony outside the $Z=50$ core is located in the $1d_{5/2}$ orbital, while from $^{123}$Sb onwards, the $0g_{7/2}$ orbital is occupied.
This change in proton orbital is evident in the spins and magnetic moments of odd-even Sb isotopes in the ground state as shown in \figref{fig:mu}a which also features the  magnetic moment of the first excited state of $^{121}$Sb with $I=7/2$ \cite{Langouche1976}.
The magnetic moments for both states are rather far away from the respective single-particle moment as already discussed in Refs.~\cite{Stone1997,Lechner2021}.

Shell-model calculations show excellent agreement with experiment when choosing appropriate $g$-factors for valence proton and neutron.
\textit{Ab initio} calculations within the VS-IMSRG framework, using bare $g$-factors, and the EM1.8/2.0 interaction also reproduce well the absolute values for Sb isotopes with $A\leq 121$, where the proton occupies the $1d_{5/2}$ orbital, although the relative trend differs from experiment. 
On the neutron-rich side ($A\geq 122$), the trend agrees nicely, but in contrast to experiment, VS-IMSRG yields magnetic moments with absolute values close to the single-particle moment of the $0g_{7/2}$ orbital.
A similar behavior had been observed for realistic shell-model calculations of $^{133}$Sb and higher-mass isotones in Ref.~\cite{Lechner2021} for which the $0g_{9/2}$ orbital below $Z=50$ had to be included for good agreement with experiment.
This had been attributed to proton excitations of the M1 type between the spin-orbit partners $0g_{7/2}$ and $0g_{9/2}$, which have a particularly strong impact on magnetic moments where the $0g_{7/2}$ orbital is involved. 
Therefore, additional \textit{ab initio} calculations were performed for $^{101,131-133}$Sb with a $^{90}$Zr core to include the proton $0g_{9/2}$ orbital, see \figref{fig:orbitals} (due to computational limitations, only isotopes next to a neutron shell closure could be calculated).
This shifts the VS-IMSRG result closer to the experimental data, see green circles in  \figref{fig:mu}a.
Note that the addition of the $0g_{9/2}$ orbital into the VS-IMSRG model space has no significant influence on the magnetic moments of the $5/2^+$ states, as verified for the $5/2^+$ state in $^{101}$Sb. 
Analogously, the inclusion of the $0h_{11/2}$ orbital did not alter the results on the magnetic and quadrupole moments.

The remaining difference between \textit{ab initio} calculations and experiment of $\approx0.5\,\mu_N$ might be caused by meson-exchange currents (MEC) presently not considered in VS-IMSRG.
Stone \textit{et al.} \cite{Stone1997} explicitly calculated the contributions from core polarization and MEC for $^{133}$Sb, albeit in a different model.
Interestingly, the obtained MEC contribution of 0.52\,$\mu_N$ to $^{133}$Sb's magnetic moment would correspond to the amount by which the VS-IMSRG misses the experimental target.
This motivates the inclusion of MEC into future VS-IMSRG calculations for Sb and other heavier systems which is work in progress.
Their explicit \textit{ab initio} application is currently limited to light nuclei~\cite{Marcucci2008,Pastore2013,Friman-Gayer2021}. 

For the shell-model calculations, effects such as core polarization and MEC are covered by effective $g$-factors, which were chosen to give the best agreement with experimental data.
Since shell-model and VS-IMSRG calculations both employ the same valence-space diagonalization, the approaches differ in their nuclear interactions within the respective valence space as well as the magnetic-moment (M1) operator and, thus, $g$-factors.
The ambition of an ab initio method is - per definition - to derive the $g$-factors from first principles.
For purely diagnostic purposes, however, it is rewarding to artificially employ the effective $g$-factor of the shell model in the VS-IMSRG calculation, see VS-IMSRG (sdg7 + eff) in  \figref{fig:mu}a.
Although the relative trends are still better reproduced by the shell model, the VS-IMSRG result based on effective $g$-factors closely matches experimental data.
This reveals that the parts of the nuclear wave functions relevant for magnetic moments are similar in both methods.
Hence, on this aspect, the EM1.8/2.0 interaction potential is almost on par with its phenomenological counterpart when looking at magnetic moments.
Thus, the comparison of $\mu$ obtained from the conventional shell model with those calculated by combining \textit{ab-initio} wave functions with the artificial use of effective $g$-factors identifies shortcomings of the M1 operator as the cause for remaining discrepancies to experiment in the proper VS-IMSRG calculation of Sb magnetic moments.
As the previous discussion suggests, one missing piece might be the lack of MEC which will hence become subject of future \textit{ab initio} work. 

\begin{figure*}
    \begin{center}
        \includegraphics[width=0.9\textwidth]{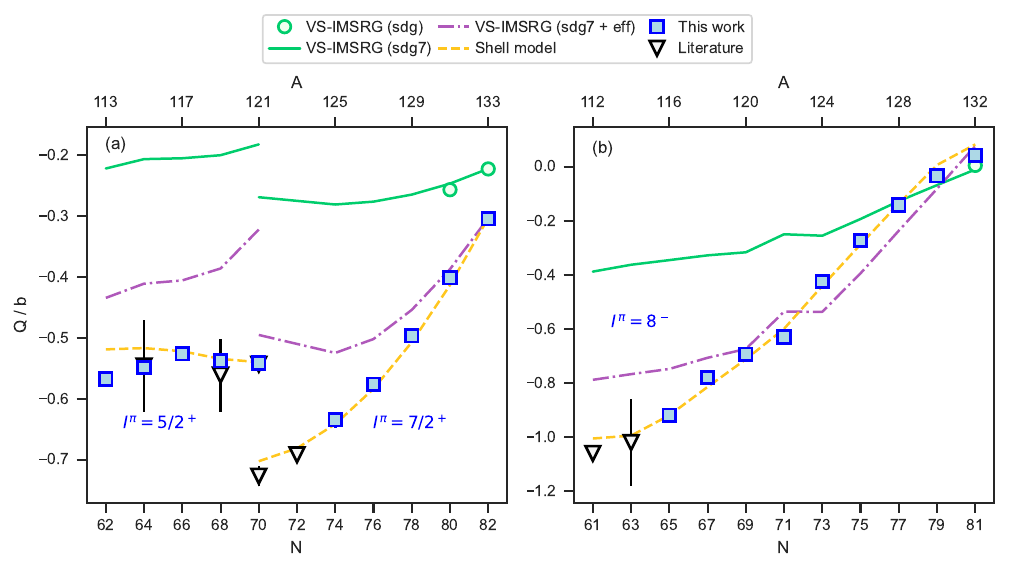}
    \end{center}
    \caption{Quadrupole moments of (a) odd-even and (b) odd-odd $^{112-133}$Sb isotopes in comparison to shell-model and \textit{ab initio} calculations within the VS-IMSRG framework and the EM1.8/2.0 nuclear interaction. 
    (sdg) includes the proton orbitals $\{0g_{9/2}, 0g_{7/2}, 1d_{5/2}, 2s_{1/2}, 1d_{3/2}\}$ and (sdg7) the same except $0g_{9/2}$, see \figref{fig:orbitals}.
    ``eff'' means that the effective charges $e_{\pi}=1.6$ and $e_{\nu}=1.05$ from the shell-model calculations were used instead of the renormalization of the operator.
    Literature values are taken from Refs.~\cite{Jackson1968,EKSTROM1974,Mahnke1982,Haiduke2006,STONE2016}.
}
    \label{fig:Q}
\end{figure*}

To further validate this perspective, we turn to our newly measured magnetic moments of $8^-$ states, plentifully found as ground or isomeric states along  the chain of odd-odd $^{112-132}$Sb isotopes.
 Here the valence neutron occupies the $0h_{11/2}$ orbital.
Despite the same spin/parity, a structural change manifests itself by a sudden discontinuity in $\mu$ between $N=69$ and $N=71$ as shown in \figref{fig:mu}b.
Once again, this is due to the proton orbital change from $1d_{5/2}$ to $0g_{7/2}$.
An overall good agreement with odd-odd magnetic moments is obtained by shell-model calculations with effective $g$-factors.
Subtle features, such as the moments of $^{120}$Sb and $^{122}$Sb, are not fully reproduced.

Similarly as for the odd-even isotopes, \textit{ab initio} calculations work fairly well on the neutron-deficient side for $8^-$ states in odd-odd isotopes, while largely underestimating the absolute values on the neutron-rich side.
Including the $0g_{9/2}$ orbital for $^{132}$Sb in the VS-IMSRG calculations brings the theoretical value very close to experiment.
This indicates that also for the odd-odd Sb isotopes with the proton in the $0g_{7/2}$ orbital, contributions of proton excitations from $0g_{9/2}$ to $0g_{7/2}$ can be significant for magnetic moments.
When utilising the shell model's effective $g$-factor in the VS-IMSRG calculation without considering the proton $0g_{9/2}$ orbital, both models reach again good agreement for neutron-rich magnetic moments, while the agreement for VS-IMSRG is not as good on the neutron-deficient side, see \figref{fig:mu}b.

\subsection{Quadrupole Moments}
The quadrupole moments of odd-even Sb isotopes also exhibit the change in proton occupation between $A=121$ and $A=123$, see \figref{fig:Q}a.
A rather constant trend is seen for the quadrupole moments on the neutron-deficient side.
Such a behavior is expected in the mid-shell region of an isotopic chain \cite{Neyens2003}.
For $A\geq 123$, a parabolic trend is observable in which the absolute values of the quadrupole moments decrease while approaching the shell closure at $N=82$.
Similar trends have been found in the Pb region \cite{Neyens2003} as well as in In~\cite{Vernon2022} and an analogy can be made for Sb:
An increasing number of neutron holes in the $h_{11/2}$ orbital interacting with the single proton causes an enhanced deformation, which explains this trend in quadrupole moments.

Quadrupole moments of odd-odd $8^-$ Sb isotopes show an almost steady decrease from a rather large oblate deformation of $Q\approx-1$\,b at $^{112}$Sb to a spherical shape at $^{132}$Sb with $Q\approx0$\,b.
A small kink in $Q$ can be observed at $N=71$, possibly again due to the proton-orbital change.

Calculations in the phenomenological shell-model yield very good agreement for quadrupole moments of odd-even and odd-odd Sb isotopes.
As already observed in earlier studies, \textit{ab initio} theory often struggles with electric quadrupole ($E2$) observables \cite{Stroberg2019,Hergert2020,Parzuchowski2017,HENDERSON2018,Heylen2021,Vernon2022,Stroberg2022}.
Therefore, the rather large underestimation of quadrupole moments of Sb isotopes in the VS-IMSRG calculations in \figref{fig:Q} is expected.
Even with the use of the shell model's effective charges, absolute values obtained via VS-IMSRG for the odd-even isotopes show a significant discrepancy.
Moreover, in the mid-shell region ($N=70-74$), the relative trends are somewhat different, too.
However, quadrupole moments of the $8^-$ in the odd-odd isotopes are fairly well reproduced with effective charges.
Compared to the shell-model results, the current VS-IMSRG framework seems to provide insufficient many-body correlations within the valence space, in addition to the missing strong $E2$ operator renormalization.

\section{Summary}

The antimony isotopic chain has been probed by collinear laser spectroscopy, revealing nuclear spins, magnetic dipole and electric quadrupole moments of $^{113-133}$Sb.
Phenomenological shell-model calculations show excellent agreement to experiment for both magnetic and quadrupole moments along the isotopic chain, once appropriate global effective $g$-factors and charges are chosen.

By exploiting the valence-space in-medium similarity renormalization group (VS-IMSRG) together with the chiral EM1.8/2.0 nuclear interaction good agreement with experiment is achieved for magnetic dipole moments of neutron-deficient Sb isotopes.
On the neutron-rich side, deviations from experiment are observed.
To reproduce the measurement data, core excitations from the $0g_{9/2}$ into the $0g_{7/2}$ orbital are shown to play an important but not necessarily sufficient role. 
Since the artificial use of the shell-model's fine-tuned effective $g$-factors in the VS-IMSRG calculations results in good agreement with shell model and experiment, the present study suggests that the relevant parts of the VS-IMSRG wave function largely match the one obtained by the phenomenological shell model.
Ongoing \textit{ab initio} advances, not yet considered in this work, are thus focused on the magnetic-moment operator by including currently neglected contributions such as meson-exchange currents.
The present analysis suggests the latter will be a key missing piece to better understand $g$-factors from first principles.
Electric quadrupole moments of Sb are underestimated by the \textit{ab initio} calculations, as already known from previous work, but relative trends are reproduced reasonably well.

The presented new experimental data motivated the first \textit{ab initio} calculations of electromagnetic moments of an entire isotopic chain above the shell closure at $Z=50$.
This marks another important step of \textit{ab initio} theory towards heavier nuclear systems.

\subsection{Acknowledgments}
We would like to thank the ISOLDE collaboration and the ISOLDE technical teams for supporting the preparation and the successful realization of the experiment.
This work was supported by the BriX Research Program No.
P7/12, FWO-Vlaanderen (Belgium), GOA 15/010 and C14/22/104 from KU Leuven, the UK Science and Technology Facilities Council grants ST/P004423/1 and ST/P004598/1, the NSF grant PHY-1068217, the BMBF Contracts No 05P18RDCIA and 05P21RDCI1, the Max-Planck Society, the Helmholtz International Center for FAIR (HIC for FAIR), the National Key R\&D Program of China (Grant No.
2022YFA1604800), the National Natural Science Foundation of China (Grant No. 12027809), the EU Horizon 2020 research and innovation programme through ENSAR2 (No.
654002), the Deutsche Forschungsgemeinschaft (DFG, German Research Foundation) -- Project-ID 279384907 -- SFB 1245,
the NSERC under grants SAPIN-2018-00027 and RGPAS-2018-522453, the Arthur B. McDonald Canadian Astroparticle Physics Research Institute,
and the European Research Council (ERC) under the European Union’s Horizon 2020 research and innovation programme (grant agreement No 101020842).
A part of VS-IMSRG computations were performed with an allocation of computing resources on Cedar at WestGrid, Compute Canada, The Digital Research Alliance of Canada and at the J\"ulich Supercomputing Center.

\bibliographystyle{elsarticle-num-names-nourl}
\bibliography{bibliography}

\end{document}